\documentclass[aps,prl,twocolumn,floatfix,showpacs]{revtex4-1}
\usepackage{epsfig,subfigure,amssymb,amscd,amsmath,graphicx,amsfonts}
\usepackage{times,longtable,braket,dsfont,bbold,txfonts}
\usepackage[usenames]{color}

\definecolor{pinegreen}{rgb}{0.0, 0.47, 0.44}

\begin{document}
 
\title{Reentrant topological phase transitions in a disordered spinless superconducting wire}

\date{\today}

\author{Maria-Theresa Rieder, Piet W. Brouwer}

\affiliation{Dahlem Center for Complex Quantum Systems and Fachbereich Physik, Freie Universit\"{a}t Berlin, Arnimallee 14, 14195 Berlin, Germany}

\author{\.{I}nan\c{c} Adagideli}

\affiliation{Faculty of Engineering and Natural Sciences, Sabanci University, Orhanli-Tuzla, Istanbul, Turkey}

\begin{abstract}
In a one-dimensional spinless $p$-wave superconductor with coherence length $\xi $, disorder induces a phase transition between a topologically nontrivial phase and a trivial insulating phase at the critical mean free path $l=\xi/2$. Here, we show that a multichannel spinless $p$-wave superconductor goes through an alternation of topologically trivial and nontrivial phases upon increasing the disorder strength, the number of phase transitions being equal to the channel number $N$. The last phase transition, from a nontrivial phase into the trivial phase, takes place at a mean free path $l = \xi/(N+1)$, parametrically smaller than the critical mean free path in one dimension. Our result is valid in the limit that the wire width $W$ is much smaller than the superconducting coherence length $\xi$.
\end{abstract}


\maketitle

In one dimension, spinless superconductors appear in two topologically distinct phases. In one of these phases, usually referred to as the ``trivial phase'' the excitation spectrum is adiabatically connected to the ionic insulator. The other phase is ``topologically nontrivial''. Topologically protected zero-energy bound states appear at junctions between the trivial and nontrivial phases \cite{Read2000,Kitaev2001}. These bound states are particle-hole symmetric and two of these combine to form a single fermionic excitation, which is why they are referred to as ``Majorana bound states'' \cite{Wilczek2009,Beenakker2011}. Interest in these systems has peaked after recent proposals to construct topological superconductors out of hybrid structures involving standard BCS superconductors and semiconductors \cite{Oreg2010,Lutchyn2010} and reports of their subsequent experimental realization \cite{Mourik2012,Das2012}. 

The Pauli principle enforces that spinless superconducting correlations are odd in momentum. In a one-dimensional setting, this means that they must be of $p$-wave type. Unlike for $s$-wave superconductors, where the Anderson theorem protects the superconducting correlations against impurity scattering \cite{Anderson1959}, backscattering by impurities suppresses $p$-wave superconducting order. In a one-dimensional wire any small amount of disorder already leads to subgap states at arbitrarily low energies, but it takes a finite amount of disorder to drive the system from the nontrivial superconducting phase into the trivial phase \cite{Motrunich2001,Brouwer2011b}. For short-range disorder with normal-state mean free path $l$, the transition between these phases takes place if \cite{Brouwer2011b}
\begin{equation}
  l = \frac{\xi}{2},
  \label{eq:1}
\end{equation}
where $\xi$ is the superconductor coherence length. Here and below we assume that the superconductivity is weak, $\xi$ much larger than Fermi wavelength $\lambda_{\rm F}$. 

The one-dimensional description applies only if the system width $W$ does not exceed the Fermi wavelength $\lambda_{\rm F}$. If $W \ge \lambda_{\rm F}$, the normal-state has $N = \mbox{int}\, (2W/\lambda_{\rm F}) > 1$ propagating channels at the Fermi level, and without disorder the topologically nontrivial phase exists if $N$ is odd, but not if $N$ is even \cite{Wimmer2010,Potter2010,Lutchyn2011,Stanescu2011,Zhou2011,Kells2012}. Numerical simulations and weakly-disordered perturbation theory indicate that the topological phases are stable against weak disorder in the multichannel case, too \cite{Rieder2012,Potter2011}. 

It is the purpose of this letter to provide an analytical theory of the effect of disorder on the topological phase in the $N$-channel $p$-wave superconductor. Our main result is that increasing the disorder strength drives the system through a sequence of $N$ topological phase transitions, taking place at
\begin{equation}
  l = \frac{n \xi}{N+1},\ \ n=1,2,\ldots,N.
  \label{eq:2}
\end{equation}
In particular, a topologically nontrivial phase persists for disorder strengths up to $l = \xi/(N+1)$, significantly larger than the critical disorder strength (\ref{eq:1}) at which the topological phase transition takes place in one dimension. Our analytical theory, as well as the precise location of the phase transitions given in Eq.\ (\ref{eq:2}), is valid in the limit of thin wires, width $W \ll \xi$. We have verified numerically that the alternation of topological phases persists for wire widths up to $W \sim \xi$.
Note that the existence of $N$ phase transitions is consistent with the known results for the weak disorder limit $l \to \infty$ (nontrivial phase if $N$ is odd, and trivial phase if $N$ is even), as well as the strong disorder limit $l \downarrow 0$ (system is in the trivial phase).

\begin{figure} [t]
\centering
 \includegraphics[width=.5\textwidth]{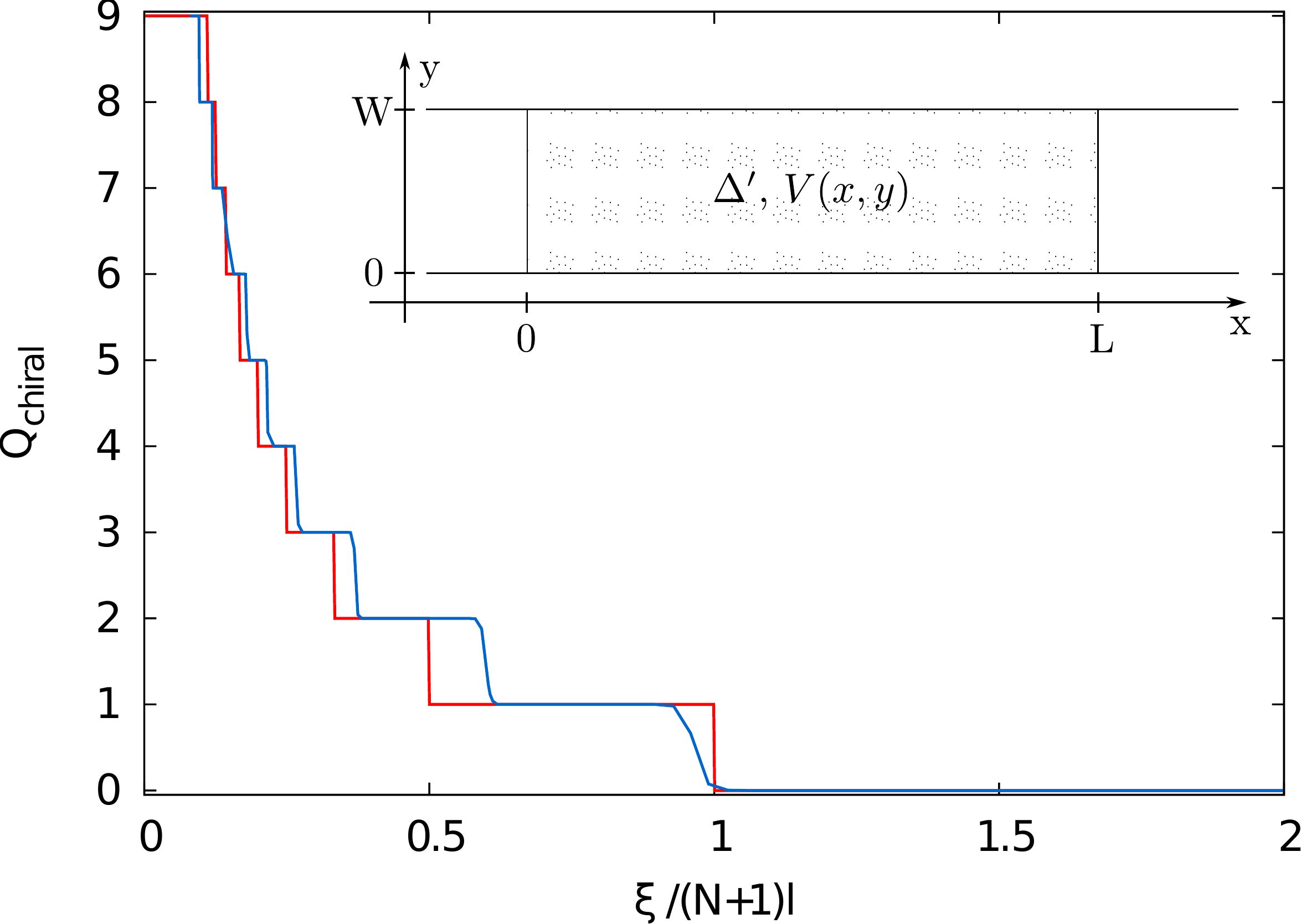}
\caption[]{(Color online.) Topological number $Q_{\rm chiral}$ for a wire with width $ 2W/\lambda_F = 9.5 $ such that the channel number is $N=9$, as a function of the ratio $\xi/l$ of disorder strength and induced superconductivity. Data shown are for a single disorder realization with $\lambda_{\rm F}/l = 0.011$ and wire length $L/l \sim 2100 $. The red curve shows the analytical prediction (\ref{eq:Qchiral3}) and the blue one the numerical data. Inset: Schematic picture of a disordered superconducting wire with two ideal normal-metal leads.}
\label{fig:BDI}
\end{figure}

For the derivation of Eq.\ (\ref{eq:2}) we consider a spinless $p$-wave superconducting wire of length $L$, width $W$, and chemical potential $ \mu > 0 $ coupled to ideal normal-metal leads at its two ends. The Bogoliubov-de Gennes Hamiltonian of the system reads
\begin{align} \label{eq:Ham_p}
  H= \left( \frac{p^2}{2m} + V(x,y) - \mu \right) \sigma_{z}+\frac{1}{2}\left\{\Delta'_x, p_{x}\right\} \sigma_{x}+\Delta'_y p_{y} \sigma_{y},
\end{align}
where $\sigma_x$, $\sigma_y$, and $\sigma_z$ are Pauli matrices in particle-hole space. The Hamiltonian (\ref{eq:Ham_p}) has particle-hole symmetry, $\sigma_x H \sigma_x = -H^*$, which places it in the (Altland-Zirnbauer) symmetry class D \cite{Altland1997}. The superconductor occupies the volume $0 < x < L$, and the two leads are at $x < 0$ and $x > L$ respectively, see Fig.\ \ref{fig:BDI} (inset). The superconducting order parameters $\Delta_x'$ and $\Delta_y'$ are nonzero for $0 < x < L$ only. The superconducting coherence length is 
\begin{equation} 
  \xi = \hbar/m \Delta_x'.
\end{equation}
Although $\Delta_x' = \Delta_y'$ for an isotropic superconductor, we have chosen to use different symbols in order to underline the very different roles of these two parameters in the calculation that follows. We assume that the superconductivity is proximity-induced, so that we can treat $\Delta_x'$ and $\Delta_y'$ as externally-imposed parameters without self-consistency requirements. 
The impurity potential has zero average and short-range fluctuations described by the Gaussian white noise correlator 
\begin{equation}
  \braket{V(x,y) V(x',y')} =\gamma \delta(x-x') \delta(y-y')
\end{equation}
and is zero in the leads. 


We determine the topological phase from the zero-energy reflection matrix $r$ of the superconducting wire \cite{Fulga2011}. In the particle-hole notation, the reflection matrix $r$ for quasiparticles incident from the left takes the form
\begin{equation}
  r = \left( \begin{array}{cc} r_{\rm ee} & r_{\rm eh} \\
  r_{\rm he} & r_{\rm hh} \end{array} \right),
\end{equation}
where particle-hole symmetry imposes that $r_{\rm hh} = r_{\rm ee}^*$ and $r_{\rm he} = r_{\rm eh}^*$ at zero energy. Following Fulga {\em et al.}, the topological phase can be calculated from the determinant $Q = \det r$ \cite{Fulga2011}: The topologically nontrivial phase has $Q = -1$, whereas the topologically trivial phase has $Q = 1$. (Note that particle-hole symmetry requires $\det r$ to be real; As no extended quasiparticle states exist in the superconductor away from the critical points, $r$ must be unitary and hence $|\det r| = 1$.)

In the thin-wire limit $W \ll \xi$ the transverse pairing $\Delta_y'$ may be treated perturbatively \cite{Kells2012}. 
Without the transverse pairing, the Bogoliubov-de Gennes Hamiltonian $H$ has an additional chiral symmetry $\sigma_y H \sigma_y = - H$ \cite{Tewari2012}, which places it in the symmetry class BD I. With the chiral symmetry, the topological superconducting phases are characterized by an integer number $Q_{\rm chiral}$. The topological quantum number $Q$ is related to $Q_{\rm chiral}$ as
\begin{equation}
  Q = (-1)^{Q_{\rm chiral}}. \label{eq:QQchiral}
\end{equation}
The absolute value $|Q_{\rm chiral}|$ can be interpreted as the number of Majorana bound states at the end of the wire, when the normal metal leads are replaced by insulating ends \cite{Fulga2011} (see also App.\ 2). The quantum number $Q_{\rm chiral}$ can be calculated from the zero-energy reflection matrix $r$ as \cite{Fulga2011}
\begin{equation}
  \label{eq:Qchiral}
  Q_{\rm chiral} = 
  -i \lim_{L \to \infty} \mbox{tr}\, r_{\rm eh}.
\end{equation}
The limit $L \to \infty$ is taken in order to ensure that the reflection matrix $r$ is unitary \cite{Fulga2011}. The chiral symmetry implies that $r_{\rm eh}$ is an antihermitian matrix, $r_{\rm eh} = - r_{\rm eh}^{\dagger}$, so that $Q_{\rm chiral}$ is real.

With the chiral symmetry present it is possible to express the zero-energy reflection matrix $r$ in terms of the system's normal-state scattering matrix at a slightly renormalized chemical potential $\tilde \mu$ \cite{Adagideli2013}. To this end, we first rotate the Hamiltonian (\ref{eq:Ham_p}) to the Majorana basis
\begin{eqnarray} \label{eq:rot_Ham}
  \tilde{H} &=& e^{-i \pi \sigma_x/4}  H  e^{i \pi \sigma_x/4} 
  \nonumber \\ &=&
- \left( \frac{p^2}{2 m} - \mu + V \right) \sigma_{y}
  +
  \Delta'_x p_{x} \sigma_{x}.
\end{eqnarray}
At zero energy, the eigenvalue equation for $\tilde{H}$ consists of two decoupled equations describing particles that are exposed to an imaginary ``gauge field'' of magnitude $\hbar/\xi$ and pointing in opposite directions for the two equations \cite{Hatano1996,Brouwer1997}. This ``gauge field'' may be transformed away by the (non-unitary) transformation
\begin{align}
  \psi(x,y) \to \tilde \psi_{\pm}(x,y) = \left\{
  \begin{array}{ll}
  \psi(x,y) & \mbox{if $x < 0$}, \\
  \psi(x,y) e^{\pm x \over \xi } & \mbox{if $0 < x < L$}, \\
  \psi(x,y) e^{\pm L/\xi} & \mbox{if $x > L$}.
  \end{array} \right.
\end{align}
The wavefunctions $\tilde \psi_{\pm}(x,y)$ satisfy the standard Schr\"odinger equation for the zero-energy wavefunction of a disordered wire,
\begin{align}
  \left( \frac{p^2}{2m} - \tilde \mu + V \right) \tilde \psi_{\pm}(x,y) = 0 \, ,
  \label{eq:psitilde}
\end{align}
where $\tilde \mu = \mu + \hbar \Delta_x'/2 \xi$. Transforming back to the basis of the original Hamiltonian (\ref{eq:Ham_p}) allows us to express the reflection matrix $r$ in terms of the reflection matrix $\tilde r$ (for particles incident from the left) and the transmission matrix $\tilde t'$ (for particles incident from the right) of the normal-state scattering problem specified by Eq.\ (\ref{eq:psitilde}),
\begin{eqnarray}
  r_{\rm ee} &=& [1 + \tilde t' \tilde t'^{\dagger} \sinh^2(L/\xi)]^{-1} \tilde r, \nonumber \\
  r_{\rm eh} &=& i \sinh(L/\xi) \cosh(L/\xi) [1 + \tilde t' \tilde t'^{\dagger} \sinh^2(L/\xi)]^{-1}
  \tilde t' \tilde t'^{\dagger}.\ \ \
\end{eqnarray}
Returning to Eq.\ (\ref{eq:Qchiral}), we find that topological number $Q_{\rm chiral}$ can be expressed as a sum over the eigenvalues $\tilde \tau_n(L)$ of $\tilde t' \tilde t'^{\dagger}$,
\begin{equation}
  Q_{\rm chiral} = \lim_{L \to \infty} \sum_{n=1}^{N}
  \frac{\sinh(L/\xi) \cosh(L/\xi) \tilde \tau_n(L)}{1 + \tilde \tau_n(L)  \sinh^2(L/\xi)}. \label{eq:Qchiral2}
\end{equation}

The asymptotic probability distribution of the eigenvalues $\tilde \tau_n$ in the limit of large $L$ is well studied in the literature \cite{Beenakker1997}. The result is best parameterized in terms of the ``Lyapunov exponents'' $\tilde \tau_n = \cosh^{-2}(x_n L)$, which are self-averaging for large $L$, with mean
\begin{equation}
  \braket{x_n} = \frac{n}{(N+1)l},\ \ n=1,2,\ldots,N,
  \label{eq:xLyapunov}
\end{equation}
and small fluctuations of order $1/\sqrt{(N+1)lL}$. The mean free path is $ l = \lambda_{\rm F} \hbar^2
v_{\rm F}^2 \alpha_N/(2 \gamma)$, with a numerical factor $ \alpha_N $ depending on the width $W$. Substituting this result into Eq.\ (\ref{eq:Qchiral2}), we find that
\begin{equation} \label{eq:Qchiral3}
  Q_{\rm chiral} = \sum_{n=1}^{N}
  \Theta \left[1 - \frac{n \xi}{(N+1)l} \right],
\end{equation}
where $\Theta(z) = 0$ if $z < 0$ and $1$ otherwise. Hence, upon increasing the disorder strength, the topological quantum number $Q_{\rm chiral}$ stepwise decreases from $Q_{\rm chiral} = N$ in the limit of zero disorder to $Q_{\rm chiral} = 0$ in the strong disorder limit. The transitions take place at the critical disorder strengths of Eq.\ (\ref{eq:2}). The topological quantum number $Q$ is given by Eq.\ (\ref{eq:QQchiral}).


When the transverse coupling proportional to $\Delta_y'$ is taken into account, the chiral symmetry is broken, and the topological quantum number $Q_{\rm chiral}$ is no longer meaningful. The quantum number $Q$ remains well defined, however. Since the effect of $\Delta_y'$ is small if $W \ll \xi$ \cite{Kells2012}, the value of $Q$ remains unchanged upon inclusion of the transverse coupling. Upon increasing the disorder strength, we therefore expect alternation between topolgical trivial ($Q=1$) and nontrivial ($Q=-1$) phases until, in the limit of strong disorder, $l < \xi/(N+1)$, the system remains in the trivial state. As long as $W \ll \xi$, the transition points should exhibit only a weak dependence on the transverse coupling $\Delta_y'$.

Alternatively (and equivalently), for a superconductor wire with hard-wall ends, the transverse coupling pairwise gaps out the $Q_{\rm chiral}$ Majorana bound states at each end of the superconducting wire, leaving behind a single Majorana state if and only if $Q_{\rm chiral}$ is odd. Since $Q_{\rm chiral}$ decreases stepwise from $N$ to zero upon increasing the disorder strength, the number of Majorana bound states at the end of the wire with $\Delta_y'$ taken into account alternates between $0$ and $1$, the transitions taking place precisely at the disorder strengths given in Eq.\ (\ref{eq:2}). Since the presence or absence of a single Majorana fermion is topologically protected, inclusion of the transverse coupling $\Delta_y' p_y$ for sufficiently small $W/\xi$ does not affect these transitions or the transition points.

For broader wires, $W \sim \xi$, the transverse coupling can not be treated perturbatively, and the results for the chiral limit $\Delta_y' \to 0$ at best have qualitative validity if the transverse coupling is included. In this respect, we note that the topological phase transitions no longer take place at weak disorder $l \gg \lambda_{\rm F}$ if $W \sim \xi$. This can be seen from Eq.\ (\ref{eq:2}) upon substituting $W \sim (N+1)\lambda_{\rm F}$, which gives $l \sim n \lambda_{\rm F}$. Equation (\ref{eq:xLyapunov}), which was essential for establishing the transition points, is derived under the assumption of weak disorder, $l \gg \lambda_{\rm F}$ \cite{Beenakker1997}, and no longer has quantitative validity if this condition is violated.

In order to further support our conclusions and to investigate the regime $W \sim \xi$, we have performed numerical simulations of the Bogoliubov-de Gennes Hamiltonian (\ref{eq:Ham_p}). We calculate the reflection matrix $r$ by concatenating short segments of length $\delta L \ll \lambda_{\rm F}$. We refer to Ref.~\onlinecite{Brouwer2011a} for a description of the numerical method. The scattering matrix of a short segment is calculated to lowest order Born approximation. For technical reasons the numerical data were obtained by varying the magnitude of the superconducting parameters $\Delta_x'$ and $\Delta_y'$ and keeping a fixed mean free path $l$.

First, we verify our analytical results in the chiral limit, $\Delta_y' = 0$. Figure \ref{fig:BDI} shows the topological number $Q_{\rm chiral}$ as a function of the ratio $\xi/l$ for a wire with $N=9$ channels. The figure clearly shows the stepwise decrease of $Q_{\rm chiral}$ upon increasing the disorder strength in comparison to the superconducting order. For coherence lengths of the order of the localization length $ (N+1)l $ the transition points closely follow the theoretical prediction (\ref{eq:2}).  We attribute the quantitative deviation of the transitions at large $ Q_{\rm chiral} $, when $ \xi \sim l $, where the relevant Lyapunov exponent $x_n$ is comparable to the inverse mean free path, to a failure of the estimate (\ref{eq:xLyapunov}) in this regime \cite{Beenakker1997}.

\begin{figure} [t]
\centering
\includegraphics[width=.5\textwidth]{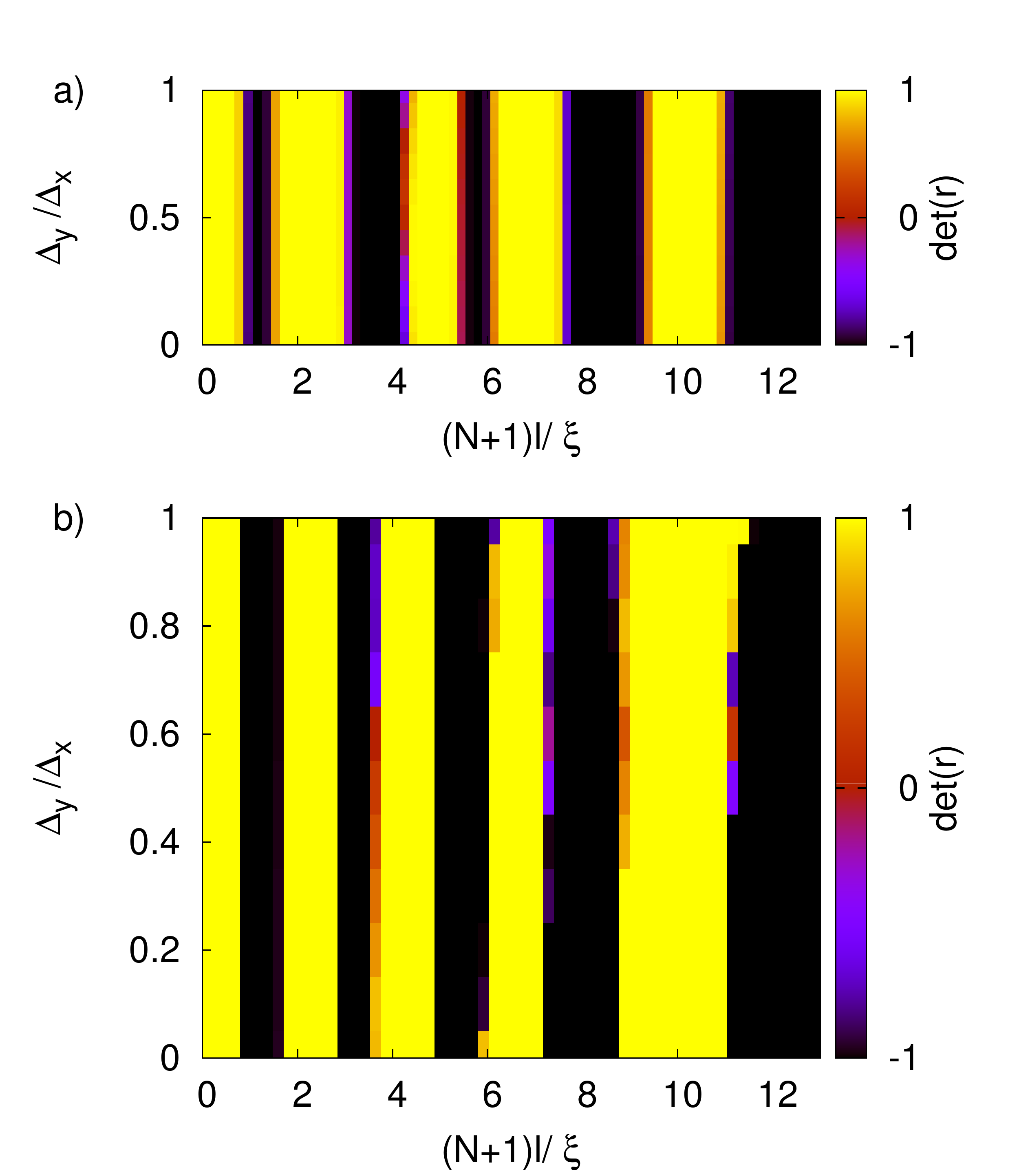}

\caption[]{(Color online.) Phase diagrams showing the topological number $Q=\det(r)$ of a wire with $ 2W/\lambda_F = 9.5 $ as a function of the ratio $l/\xi$ and the transverse coupling $\Delta_y'$ in a spinless superconducting wire witih $N=9$ channels in the limit of weak and strong disorder (panel \textbf{a)} and \textbf{b)}, resp.). The disorder strength is $\lambda_{\rm F}/l = 0.011 $ (panel \textbf{a)}) and $\lambda_{\rm F}/l =0.43$ (panel \textbf{b)}). The data were obtained for a wire length $L/l = 150 $ (panel \textbf{a)}) and $L/l \sim 770 $ (panel \textbf{b)}).}
\label{fig:Biene_Maja}
\end{figure}

The effect of the transverse coupling on the phase transitions is shown in Fig.\ \ref{fig:Biene_Maja}. Both panels of Fig.\ \ref{fig:Biene_Maja} show the value of $Q = \det r$ as a function of $(N+1)l/\xi$ and of $\Delta_y'/\Delta_x'$ at a fixed realization of the random potential $V$ ({\em i.e.}, at a fixed value of the mean free path $l$). The top panel of this figure shows representative numerial data for a weakly disordered wire ($\lambda_{\rm F}/l = 0.011$), where all transitions take place within the limit $W \ll \xi$. As expected, the sequence of topologically trivial and nontrivial phases does not significantly depend on the transverse coupling $\Delta_y'$ in this case. The bottom panel shows data for strong disorder ($\lambda_{\rm F}/l =0.43$), where the condition $W \ll \xi$ is no longer satisfied for small values of $\xi/l$. The disorder strength is chosen such that the condition $ W \simeq \xi $ is met roughly at the 6th transition. For the necessarily finite wire lengths $L$ in the numerical 
simulations, finite-size effects lead to a blurring of the topological phase transitions. The occurence of values of $\det r$ different from $-1$ or $1$ signals a breakdown of the insulating behavior of the superconductor. This behavior is consistent with Ref.~\cite{Brouwer2003}, where it was found that for large $N$ a spinless superconducting wire enters a quasi-critical region with algebraic instead of exponentially decaying transmission \cite{Brouwer2000}, which persists up to wire lengths $L$ much larger than the normal-state localization length and out of range of our numerical simulations. It is also consistent with the approach of the two-dimensional limit, in which the one-dimensional thermal insulator transitions into a two-dimensional thermal metal \cite{Medvedyeva2010}.
 
In conclusion, we investigated the effect of disorder on the topological phase in a multichannel p-wave superconducting wire. From an analytical study in the limit of thin wires $ W \ll \xi $, we derived a series of topological phases transitions in which the system alternates between trivial and nontrivial phases. A numerical analysis shows that this holds true also for thicker wire $ W \lesssim \xi $

We gratefully acknowledge discussions with Michael Wimmer and Falko Pientka. This work is supported by the Alexander von Humboldt Foundation and by the funds of the Erdal Inonu chair.

\paragraph{Appendix 1: Full scattering matrix.}

In the main text, we used the mapping to the normal scattering problem (\ref{eq:psitilde}) to compute the left reflection matrix of the superconductor. The same mapping can be used to compute the full scattering matrix. We here summarize the main resuls.

The scattering matrix of the normal metal wire may be written in the polar demoposition
\begin{eqnarray}
 \widetilde{S} &=& \begin{pmatrix} \tilde r & \tilde t' \\ \tilde t & \tilde r'
  \end{pmatrix} \nonumber \\ &=&
\begin{pmatrix} U & 0 \\ 0 & V \end{pmatrix}
 \begin{pmatrix} -\sqrt{1- \tilde \tau} & \sqrt{\tilde \tau} \\ \sqrt{\tilde \tau} & \sqrt{1- \tilde \tau} \end{pmatrix}
 \begin{pmatrix} U^{\rm T} & 0 \\ 0 & V^{\rm T} \end{pmatrix} \, ,
\end{eqnarray} 
where $ U $ and $V$ are unitary matrices and $\tilde \tau = \text{diag} \left( \tilde \tau_{1}, \tilde \tau_{2},\ldots,\tilde \tau_{n} \right)$. For the scattering matrix $S$ of the superconductor we then find
\begin{eqnarray} \label{eq:scatmat_p}
 S &=& 
\begin{pmatrix}
     r & t' \\ t & r' 
    \end{pmatrix}
  \\ \nonumber &=&
 \begin{pmatrix}
      U & 0 & 0 & 0 \\ 0 & U^{\ast} & 0 & 0 \\ 0 & 0 & V & 0 \\ 0 & 0 & 0 & V^{\ast} 
     \end{pmatrix}
     \begin{pmatrix}
       - \rho & \bar \rho & \tau & \bar \tau   \\ 
       -\bar \rho & -\rho & -\bar \tau  & \tau \\ 
      \tau  &\bar \tau &\rho & -\bar \rho  \\ 
       -\bar \tau & \tau & \bar \rho & \rho
     \end{pmatrix}
     \begin{pmatrix}
     U^{\rm T} & 0 & 0 & 0 \\ 0 & U^{\dagger} & 0 & 0 \\ 0 & 0 & V^{\rm T} & 0 \\ 0 & 0 & 0 &  V^{\dagger} 
     \end{pmatrix},
\end{eqnarray}
where we abbreviated
\begin{align}
  \rho &= \left[ 1 + \tilde \tau  z_{-}^{2} \right]^{-1} \sqrt{1- \tilde \tau} \nonumber \\
  \bar \rho &= i \left[ 1 + \tilde \tau z_{-}^{2} \right]^{-1} \tilde z_{+} z_{-} \tau\nonumber \\
  \tau &= \left[ 1 + \tilde \tau  z_{-}^{2} \right]^{-1} z_{+} \sqrt{\tilde \tau}   \nonumber \\
  \bar \tau &= i \left[ 1 + \tilde \tau  z_{-}^{2} \right]^{-1}  z_{-} \sqrt{\tilde \tau (1-\tilde \tau)},
\end{align}
with $ z_{+} = \cosh(L/\xi) $ and $ z_{-} = \sinh(L/\xi)$. 

\paragraph{Appendix 2: Number of Majorana fermions.}

To connect the topological quantum number $Q_{\rm chiral}$ to the number of Majorana fermions for a superconducting wire with hard-wall ends, we recall that the number of Majorana fermions is equal to the dimension of the nullspace of $1 + r$. Making use of our explicit solution (\ref{eq:scatmat_p}) for the reflection matrix $r$ in terms of diagonal matrices $\rho$ and $\bar \rho$, we see that the elements of the diagonal matrix $\rho$ are $0$ or $1$ in the limit $L \to \infty$, and that $\bar \rho = i (1-\rho)$. We may then factorize $1 + r$ as
\begin{eqnarray}
  1 + r &=&
 U \begin{pmatrix}  U^{\dagger} U^{\ast} -\rho & i(1-\rho) \\ -i(1-\rho) &  U^{\rm T} U  - \rho \end{pmatrix} U^{\rm T} \nonumber \\ &=&
  U
\begin{pmatrix}  U^{\dagger} & -\rho \\ -i U^{T} &   i \rho \end{pmatrix}
\begin{pmatrix}  U^{\ast} & iU \\ \rho  &   i \rho \end{pmatrix} U^{\rm T}.
\end{eqnarray}
From this expression one immediately concludes that the dimension of the nullspace of $1 + r$ is the number of unit eigenvalues of $-i \bar \rho$, which is equal to $Q_{\rm chiral}$ by Eq.\ (\ref{eq:Qchiral}).

\paragraph{Appendix 3: Mean free path}
The mean free path in the normal state is calculated from the reflection matrix for a short segment of length $dL $ as $ \text{tr}\left( r_{\rm ee} r_{\rm ee}^{\dagger} \right) = N dL/l $, which is calculated in the first order Born approximation. The expression for $l$ given in the main text contains a factor $\alpha_N$ that explicitly reads
\begin{align}
 \alpha_N = N &\left[\frac{3}{2} \sum\limits_{n=1}^{N} \left( \left(Wk_F/\pi \right)^{2}-n^{2} \right)^{-1} \right. \nonumber \\
 &\left. + 2\sum\limits_{n<m=1}^{N} \sqrt{(\left(Wk_F/\pi \right)^{2}-n^{2})^{-1}(\left(Wk_F/\pi \right)^{2}-n^{2})^{-1}} \right]^{-1} \, .
\end{align}

\end{document}